\documentstyle[twoside,fleqn,sprocl,epsfig]{article}
\newcommand{\AmS}{{\protect\the\textfont2
  A\kern-.1667em\lower.5ex\hbox{M}\kern-.125emS}}

\begin{document}

\title{THE HIGH DENSITY PARTON DYNAMICS FROM EIKONAL AND DIPOLE PICTURES}

\author{ M. B. Gay  Ducati and V. P. Gon\c{c}alves }
\address{Instituto de F\'{\i}sica, Univ. Federal do Rio Grande do Sul \\
 Caixa Postal 15051, 91501-970 Porto Alegre, RS, BRAZIL}

\maketitle


\begin{abstract}
\small{In this contribution we discuss  distinct approaches for the high density QCD (hdQCD)  and the demonstration of the equivalence between the AGL approach and the K  approach is presented.  
  Our conclusion is that the AGL equation is a good candidate for an unitarized evolution equation at high densities in the DLA limit.}

\end{abstract}


\section{Introduction}

The behavior of the cross sections in the high energy limit ($s \rightarrow \infty$) and fixed momentum transfer is expected to be described by the BFKL equation. The simplest process where  this equation applies  is the high energy scattering between two heavy quark-antiquark states, {\it i.e.} the onium-onium  scattering.
This process was studied in the dipole picture \cite{mueller},  where the 
heavy quark-antiquark pair and the 
soft gluons in the limit of large number of colors  $N_c$  are viewed as a collection of color 
dipoles. 
One of the main characteristics of the BFKL equation is that it predicts 
very high density of partons  in the small $x$ region. At sufficient small values of $x$, one enters in the  regime of the high density QCD where partons from neighbouring ladders overlap spatially and new dynamical effects  associated with the unitarity corrections are expected to stop further growth of the parton densities. The understanding and analytic description of the unitarity corrections and consequently of the high density QCD is currently one of the great  challenges.

 At the moment, there are three distinct perturbative approaches for the dynamics at high densities (See Fig. 1): 
\begin{itemize}

\item the Ayala, Gay Ducati and Levin (AGL) approach \cite{ayala1}, which considers the  multiple pomeron exchange in the double logarithmic approximation (DLA), considering as basic degree of freedom the usual partons (quarks and gluons).
 The starting point of this approach is the proof of the Glauber formula in QCD \cite{muegla}, which considers only the interaction  of the fastest partons  with the target.
In \cite{ayala1}, a generalized equation which takes into account the interaction 
of all partons in a parton cascade with the target in the DLA limit  was proposed. 
The main properties of the  generalized equation  are that it  resums all multiple pomeron exchanges in the DLA limit
and  its  assintotic solution is given by $xG \propto Q^2\,R^2\,ln\,(1/x)$, where $R$ is the size of the target,  {\it i.e.} differently from the GLR equation, it does not  predict saturation of  the gluon distribution  in the very small $x$ limit.

\item the McLerran-Venugopalan {\it et al.} (MV) approach \cite{mcl,jamal}, which is based on  effective Lagrangian formalism for the low $x$ DIS  and the Wilson renormalization group. The basic degree of freedom is the gluonic field.
An unitarized evolution equation was proposed by Jalilian-Marian {\it et al.} \cite{jamal} using the MV approach. These authors have derived a general evolution equation for the gluon distribution  in the limit of large parton densities and leading logarithmic approximation, considering  a very large nucleus. In the general case the evolution equation  is a very complicated equation, which does not allow to obtain analytical solutions. Recently, these authors have considered the DLA limit on their result \cite{jamal1} and have shown 
that the evolution equation reduces to an equation with a functional form similar, but not identical, to the AGL equation.

\item the Kovchegov  (K) approach \cite{kov}, which is based in the resumation of the multiple pomeron exchange in the leading logarithmic $1/x$ approximation using the dipole picture, where the basic degree of freedom are $q \overline{q}$ dipoles. The author has obtained an evolution equation for the cross section of the  $q \overline{q}$ pair with a nucleus $N(\vec{x_{01}}, \vec{b_0}, y)$ considering the multiple scattering of the dipoles which unitarizes the BFKL Pomeron.

\end{itemize}

In contrast with the MV {\it et al.} and K approaches, in the AGL approach a comprehensive phenomenological analysis   
 of the $ep$ \cite{ayala3} and $eA$ processes \cite{f2a,df2a} exists. 
The analysis of the behavior of distinct observables  for the HERA kinematical region using the Glauber-Mueller formula was presented in  \cite{ayala3}. In this kinematical region the solutions from the AGL equation and the Glauber-Mueller formula approximately coincide. 
The results from these analysis agree with the recent HERA data and allows to make some predictions which will be tested in a near future. Our main conclusion was that the unitarity corrections cannot be disregarded in the HERA kinematical region.

All these approaches reproduce the DLA limit of the DGLAP evolution equations in the DLA limit of PQCD and the GLR equation as the first order unitarity correction. The main open questions are: Since at high densities higher orders cannot be disregarded,  what  is the correct framework  to treat these  unitarity corrections? There is a common limit between these approaches?  

In this contribution we present  the main steps of the demonstration of  equivalence  between the K equation and    the AGL equation in the DLA limit \cite{agl}.  This result motivates to use the 
AGL approach,  which considers the medium effects associated with the high density present at small values of $x$ and/or nuclear collisions at high energies, as the dynamics at dense systems.

\begin{figure}
\centerline{\psfig{file=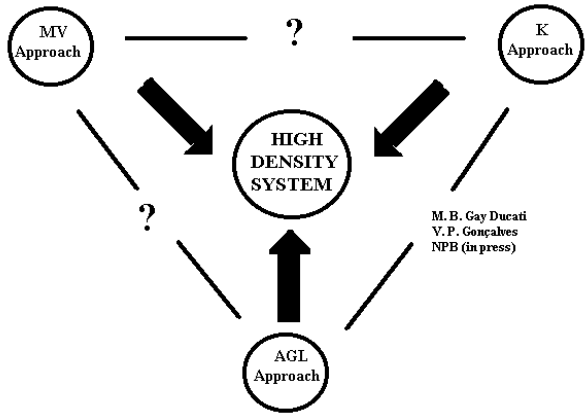,width=100mm}} 
\caption{ The present map of the approaches for the high density QCD.}
\label{fig1}
\end{figure}

\section{The Unitarity Corrections}

We start considering  the interaction between a virtual colorless hard probe and the nucleus via a gluon pair component of the virtual probe.  The cross section for this process is written as
\begin{eqnarray}
\sigma^{G^*A}=  \int_0^1 dz \int \frac{d^2r_t}{\pi} 
 |\Psi_t^{G^*}(Q^2,r_t,x,z)|^2 \sigma^{gg+A}(z,r_t^2)\,\,,
\label{sig1}
\end{eqnarray}
where $G^*$ is the  virtual colorless hard probe with virtuality  $Q^2$, $r_t$ is the transverse separation of the pair, $z$ is the fraction of energy carried by the gluon and $\Psi_t^{G^*}$ is the wave function of the transverse polarized gluon in the virtual probe. Furthermore, $\sigma^{gg+A}(z,r_t^2)$ is the cross section of the interaction of the $gg$ pair with the  nucleus.
 Considering that $\sigma^{gg + A} = (C_A/C_F) 
\sigma^{q \overline{q}}$ we have that the cross section for the interaction 
between the virtual probe $G^*$ and the nucleus via the $q \overline{q}$ component
of the virtual probe is given by
\begin{eqnarray}
\sigma^{G^*A}= \frac{C_A}{C_F} \int_0^1 dz \int \frac{d^2r_t}{\pi} 
 |\Psi_t^{G^*}(Q^2,r_t,x,z)|^2 \sigma^{q\overline{q}+A}(z,r_t^2)\,\,.
\label{sig2}
\end{eqnarray}
To estimate the unitarity corrections we have to take into account the rescatterings of the gluon pair inside the nucleus. Using the Glauber-Mueller approach, which  considers the interations of the fastest partons with the target, we get 
\begin{eqnarray}
\sigma^{G^*A}= \frac{C_A}{C_F}  \int_0^1 dz \int \frac{d^2r_t}{\pi} 
\int \frac{d^2b_t}{\pi} |\Psi_t^{G^*}(Q^2,r_t,z)|^2 \,2\,
[1 - e^{-\frac{1}{2}\sigma_N^{q\overline{q}}(x^{\prime}
,\frac{4}{r_t^2})S(b_t)}]\,\,, \nonumber
\label{sig3}
\end{eqnarray}
where $x^{\prime} = x/(z\,r_t^2\,Q^2)$ ($x$ is the Bjorken variable), $ b_t$ is the impact parameter,  
$S(b_t) = (A/ \pi R^2) e^{-\frac{b_t^2}{R^2}}$ is the gaussian profile 
function  and $\sigma_N^{q\overline{q}}$ is the cross section of the 
interaction of the $q\overline{q}$ pair with the  nucleons inside  the 
nucleus. It was   shown \cite{plb} that $ \sigma_N^{q\overline{q}} 
= \frac{C_F}{C_A} (3 \alpha_s(\frac{4}{r_t^2})/4)\,\pi^2\,r_t^2\,
 xG_N(x,\frac{4}{r_t^2})$, where  $xG_N(x,\frac{4}{r_t^2})$ is the nucleon gluon 
 distribution.

The  relation $\sigma^{G^*A}(x,Q^2) = (4\pi^2 \alpha_s/Q^2)xG_A(x,Q^2)$ is valid for a virtual probe $G^*$ with virtuality $Q^2$.  
Consequently, using  the expression of the squared  wavefunction 
 we obtain that   
the Glauber-Mueller formula for the interaction of the $q\overline{q}$ pair with the 
nucleus is written as
\begin{eqnarray}
xG_A(x,Q^2) = \frac{4}{\pi^2}  \frac{C_A}{C_F} \int_x^1 
\frac{dx^{\prime}}{x^{\prime}} \int_{\frac{4}{Q^2}}^{\infty} \frac{d^2r_t}{\pi r_t^4}
\int \frac{d^2b_t}{\pi} \,2\, [1 - e^{- \frac{1}{2} \sigma_N^{q \overline{q}} (x^{\prime}, \frac{4}{r_t^2}) S(b_t)}]\,\,. \nonumber
\label{gm}
\end{eqnarray}

The AGL equation can be obtained directly from the above equation  
differentiating this formula with respect to $y = ln \,1/x$ and 
$\epsilon = ln \, Q^2/\Lambda_{QCD}^2$. Therefore the AGL for the interaction of 
$q\overline{q}$ dipole is given by 
\begin{eqnarray}
\partial^2_{y \epsilon} xG_A(x,Q^2) = 
\frac{2\,Q^2}{\pi^2} \frac{C_A}{C_F}  
\int \frac{d^2b_t}{\pi}  \,[1 - e^{-\frac{1}{2}\sigma_N^{q\overline{q}}(x
,Q^2)S(b_t)}]\,\,, 
\label{agl}
\end{eqnarray}
where the dependence of  $\sigma_N^{q\overline{q}}$ in the  virtuality of the   
virtual probe results from the derivative. The nonperturbative effects 
coming from the large distances are absorbed in the boundary and initial 
conditions. This equation is valid in the double logarithmic approximation (DLA).

Considering a central collision ($b=0$) and that the  transverse cross-sectional area of the nucleus is $S_{\bot} = \pi  R^2$ and that $S(0) = A/(\pi R^2)$, the AGL equation for $b=0$ is obtained as
\begin{eqnarray}
\partial^2_{y \epsilon}xG_A(x,Q^2) = \frac{\cal{C}}{\pi^3}
   Q^2[1 - e^{-\frac{2\alpha_s \pi^2 }{N_c S_{\bot}} \frac{1}{Q^2
} xG_A(x,Q^2)}]\,, 
\label{aglkov}
\end{eqnarray}
where ${\cal{C}} = N_c \,C_F \, S_{\bot}$. This equation takes into account that each parton in the parton cascade interacts with the nucleons within the nucleus (Glauber multiple scattering).

The GLR equation can be obtained directly from the AGL equation. If we expand the right hand side of this equation to the second order in $xG_A$ we obtain
\begin{eqnarray}
\partial^2_{y \epsilon} xG_A(x,Q^2)  = \frac{\alpha_s N_c}{\pi} \,xG_A(x,Q^2) 
- \frac{\alpha_s^2 \pi}{S_{\bot}} \frac{1}{Q^2} [xG_A(x,Q^2)]^2 \,\,, 
\label{glr}
\end{eqnarray}
which is the GLR equation for a cylindrical nucleus case. Moreover, if the unitarity corrections are small, only the first order in $xG_A$ contributes. In this limit the AGL equation  matches with the DGLAP evolution equation in the DLA limit.

Considering the multiple pomeron exchanges, Kovchegov \cite{kov} has obtained an evolution equation for the total cross section  of the  $q\overline{q}$ pair with a transverse size $x_{01}$   interacting with the nucleus $N(\vec{x_{01}}, \vec{b_0}, y)$ in the leading logarithmic approximation. The K equation was obtained considering the scattering of a virtual photon with a nucleus. The physical picture for this interaction is the same as the Glauber-Mueller approach. The incoming virtual photon generates a  $q\overline{q}$ pair which develops a cascade of gluons, which then scatters on the nucleus. In the large $N_c$ limit the gluon can be represented as a $q\overline{q}$ pair. Therefore, in this limit  and in the leading logarithmic approximation, the cascade of gluons can be interpreted as a dipole cascade, where each dipole in the cascade interacts with several nucleons within  the nucleus. Therefore, as the  K equation  and the AGL equation, although with distinct basic objects, resums the multiple scatterings of its degree of freedom, we expect that both coincide in a common limit.
In the double logarithmic limit the K equation reduces to
\begin{eqnarray}
\partial^2_{y \epsilon^{\prime}} N(\vec{x_{01}}, \vec{b_0}, y) = 
\frac{\alpha_s C_F}{\pi}\,[2  -  N(\vec{x_{01}}, \vec{b_0}, y)] N(\vec{x_{01}}, \vec{b_0}, y)\,, 
\label{kovdif}
\end{eqnarray}
where  $\epsilon^{\prime} = ln \,(1/x_{01}^2 \Lambda_{QCD}^2)$. We denote the above expression K (DLA) equation.

In \cite{kov} the connection between the quantity  $N(\vec{x_{01}}, \vec{b_0}, y)$ and the nuclear gluon distribution was discussed.  As there is some freedom in the definition of the gluon distribution, the choice for the connection between the two functions was arbitrary. In \cite{agl} we have used the result obtained from  the nuclear structure function to establish the relation between these functions. The nuclear structure function in the target rest frame is given by 
\begin{eqnarray}
F_2^A(x,Q^2) = \frac{Q^2}{4 \pi \alpha_{em}} \int dz \int \frac{d^2r_t}{\pi} |\Psi(z,r_t)|^2 \, \sigma^{q\overline{q} + A}(z,r_t)\,\,.
\label{f2target}
\end{eqnarray}
In the Glauber-Mueller approach the cross section for the interaction of the $q\overline{q}$ pair with the  nucleus
$\sigma^{q\overline{q} + A}(z,r_t)$, in a central collision ($b=0$), is given by $\sigma^{q\overline{q} + A}(b=0) = R^2
\,2\,\{1 - e^{-\frac{1}{2}\sigma_N^{q\overline{q}}S(0)}\}\,\,.$
This allows to estimate the unitarity corrections to the nuclear structure function for central collisions ($b=0$) in the DLA limit.

Comparing with the expression of the nuclear structure function in the dipole picture proposed in \cite{kov}, we obtain that
  the total cross section of the $q\overline{q}$ pair interacting with the nucleus, $N(\vec{x_{01}}, \vec{b_0} = 0, y)$, is given by
\begin{eqnarray}
N(\vec{x_{01}}, \vec{b_0} , y) = \,2\,\{1 - e^{-\frac{ \alpha_s C_F \pi^2}{N_c^2 S_{\bot}} x_{01}^2 AxG(x,1/x_{01}^2)}\}\,. \nonumber
\label{ene}
\end{eqnarray}
Substituting the  above expression in the K (DLA) equation and using $x_{01} \approx 2/Q$, the AGL equation is a straithforward consequence \cite{agl}.

\section{Conclusion}
We demonstrate in this contribution that the AGL equation can be obtained from the dipole picture. This result shows that the AGL equation is a good candidate for the unitarized evolution equation at small $x$ in the DLA limit, supported by two different frameworks describing high density phenomena. However, we still need to obtain the connection between the AGL  approach and the  MV approach to establish the limits of the distinct approaches and  the theory of  hdQCD (see Fig. 1).  Recently, 
 Jalilian-Marian {\it et al.} \cite{jamal} have considered the DLA limit on their result \cite{jamal1} and have shown 
that the evolution equation reduces to an equation with a functional form similar, but not identical, to the AGL equation. We believe that a more detailed analysis of the approximations used in both equations will allow to  demonstrate the equivalence in a common limit.

\end{document}